\documentstyle[aps,prl,multicol,graphics,epsfig,enumerate,amstex]{revtex}

\begin{document}
\draft
\title{Quantum Conductance in Semimetallic Bismuth Nanocontacts}
\author{J.G. Rodrigo, A. Garc\'{\i}a-Mart\'{\i}n\cite{adress}, J.J. S\'aenz and
S.Vieira}
\address{Instituto Universitario de Ciencia de Materiales ``Nicol\'as Cabrera'' \\
and \\
Laboratorio de Bajas Temperaturas,\\
Departamento de F\'{\i}sica de la Materia Condensada, \\
Universidad Aut\'onoma de Madrid, 28049 Madrid, Spain}
\maketitle

\begin{abstract}
Electronic transport properties of bismuth nanocontacts are analyzed by
means of a low temperature scanning tunneling microscope. The subquantum
steps observed in the conductance versus elongation curves give evidence
of atomic rearrangements in the contact. The underlying quantum nature of
the conductance reveals itself through peaks in the conductance
histograms. The shape of the conductance curves at 77 K is well described
by a simple gliding mechanism for the contact evolution during elongation.
The strikingly different behaviour at 4 K suggests a charge carrier
transition  from light to heavy ones as the contact cross section becomes
sufficiently small.
\end{abstract}

\pacs{PACS numbers: 61.16.Ch, 73.50.-h, 73.23.Ad, 73.20.Dx}

\begin{multicols}{2}
In the last few years, several techniques have been extensively used to
study the conductance of nanometer-scale contacts with variable
cross sections \cite{Muller,Agrait,Costa1,Nanowires}. Most of the previous
works have been focused on metallic contacts, for which the Fermi
wavelength  $\lambda _F$ is very close to the dimensions of the atom.  
The elongation of the contact
produces discrete changes in the cross section (i.e. in the number of
atoms) resulting in abrupt jumps in the experimental $G-Z$ curves
(conductance vs. elongation of the contact). These jumps are of the order
of the conductance quantum $G_0=2e^2/h$ since the addition of one atom to
the contact is accompanied by the opening of one or few conduction
channels depending on the metal\cite{channels}. This interplay between
electronic and mechanical properties in metallic contacts 
\cite{Landman,Agrait2} makes it difficult to observe ``pure'' quantum
effects on the conductance.

On the other hand, semimetallic nanocontacts could provide a unique
laboratory to investigate quantum conductance phenomena in
three-dimensional wires. The high value of $\lambda _F$ in these systems 
requires large contact areas in order to have total transmissions of the
order of a conductance quantum. As a consequence, it will  be possible to
distinguish between atomic rearrangements and quantum effects. However,
earlier experimental results did not clarify the situation. The first
experiments \cite{Krans-Sb} on semimetallic Sb contacts showed several
sharp jumps in the $G-Z$ curves for conductances much lower than the
quantum unit, proving that, indeed, many atoms are necessary in order to
obtain a value of the conductance equal to $G_0$. Nevertheless the
conductance quantization features were absent probably due to a strong
backscattering \cite{Krans-Sb} or to the large angular opening
\cite{Torres94} of the contact. More recently, conductance
experiments\cite{Olin} on Bi nanowires at 4K presented a very different
behaviour showing, in some situations, long conductance plateaus close to
integer multiples of $G_0$. After statistical averaging of many
consecutive curves, the histogram of conductance values showed the
existence of two peaks close to $G_0$ and $2G_0$. However, the shape of
the histogram exhibited strong deviations from those reported in metallic
contacts.
\begin{figure}[]
\epsfig{file=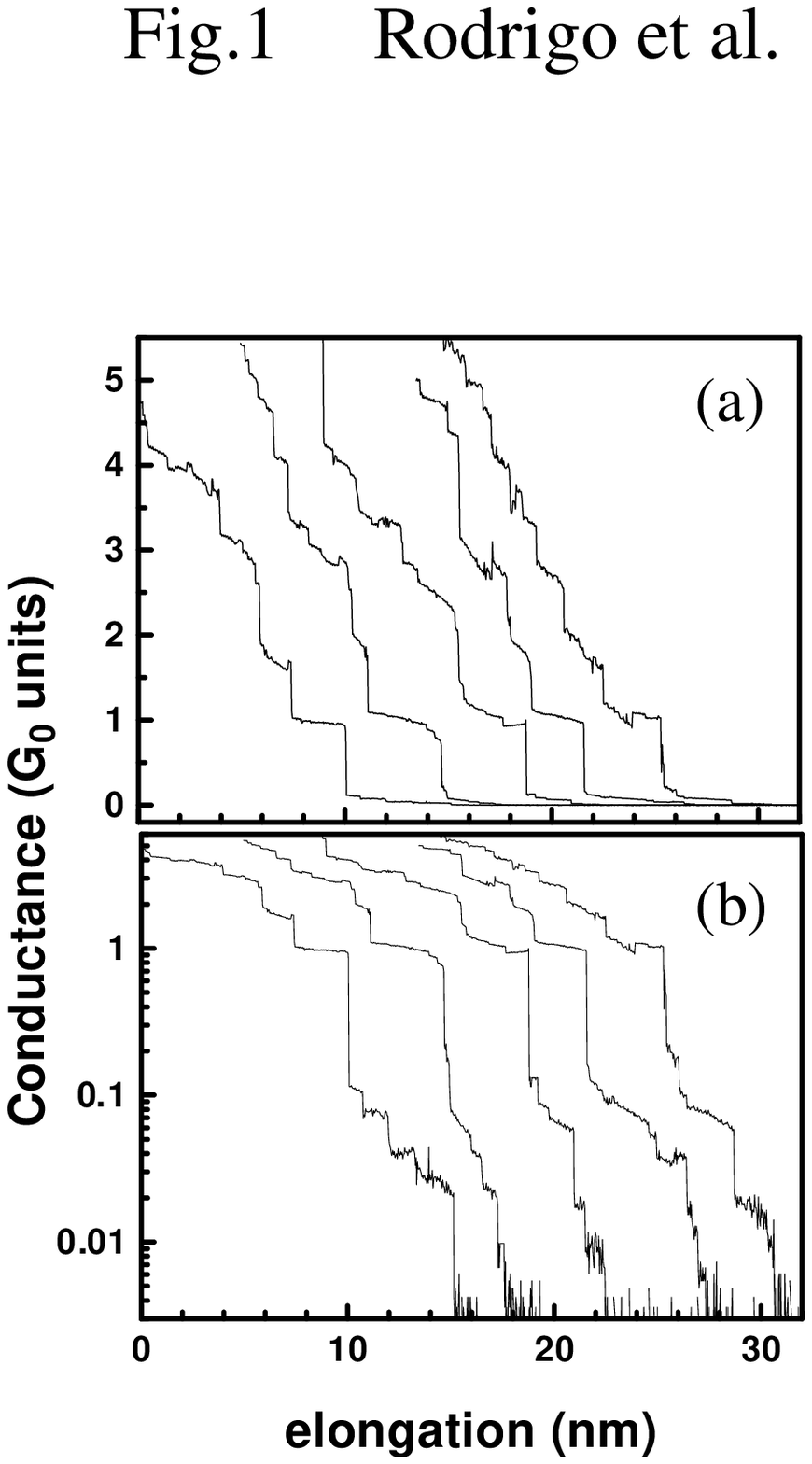,width=8cm,clip=}
\caption{ Typical $G-Z$ curves corresponding to elongation-contraction
cycles measured at 77 K: (a) linear scale, (b)logarithmic scale. Curves are
shifted horizontally for clarity.}
\label{fig1}
\end{figure}

In order to shine some light on the physics of semimetallic contacts, we
have performed scanning tunneling microscopy (STM) experiments on Bi
nanocontacts at different temperatures. First, we analyze in detail the
opening of the first conduction channel, which is a particularly difficult
task in metallic contacts, where the transition from the tunneling regime
into contact is governed by the well known jump-to-contact
phenomenon\cite{Pethica}. The peculiar electronic structure of
bismuth\cite{Bitheory,PRL2001} enables a selection of the carrier type, depending
on the dimensions of the contact and temperature\cite{Zhang}. This effect
will be studied at 77K and 4.2K for contacts with typical conductances in
the range $0<G<2G_0$. The coexistence of conductance quantization effects
and atomic rearrangements will be studied through conductance histograms
of the $G-Z$ curves. Due to the fact that even for low conductance values
many atoms will be forming the contact, the evolution of the $G-Z$ curves
will be analyzed in terms of a simple gliding mechanism that describes the
variation of the contact shape during the elongation process.

\begin{figure}[]
\epsfig{file=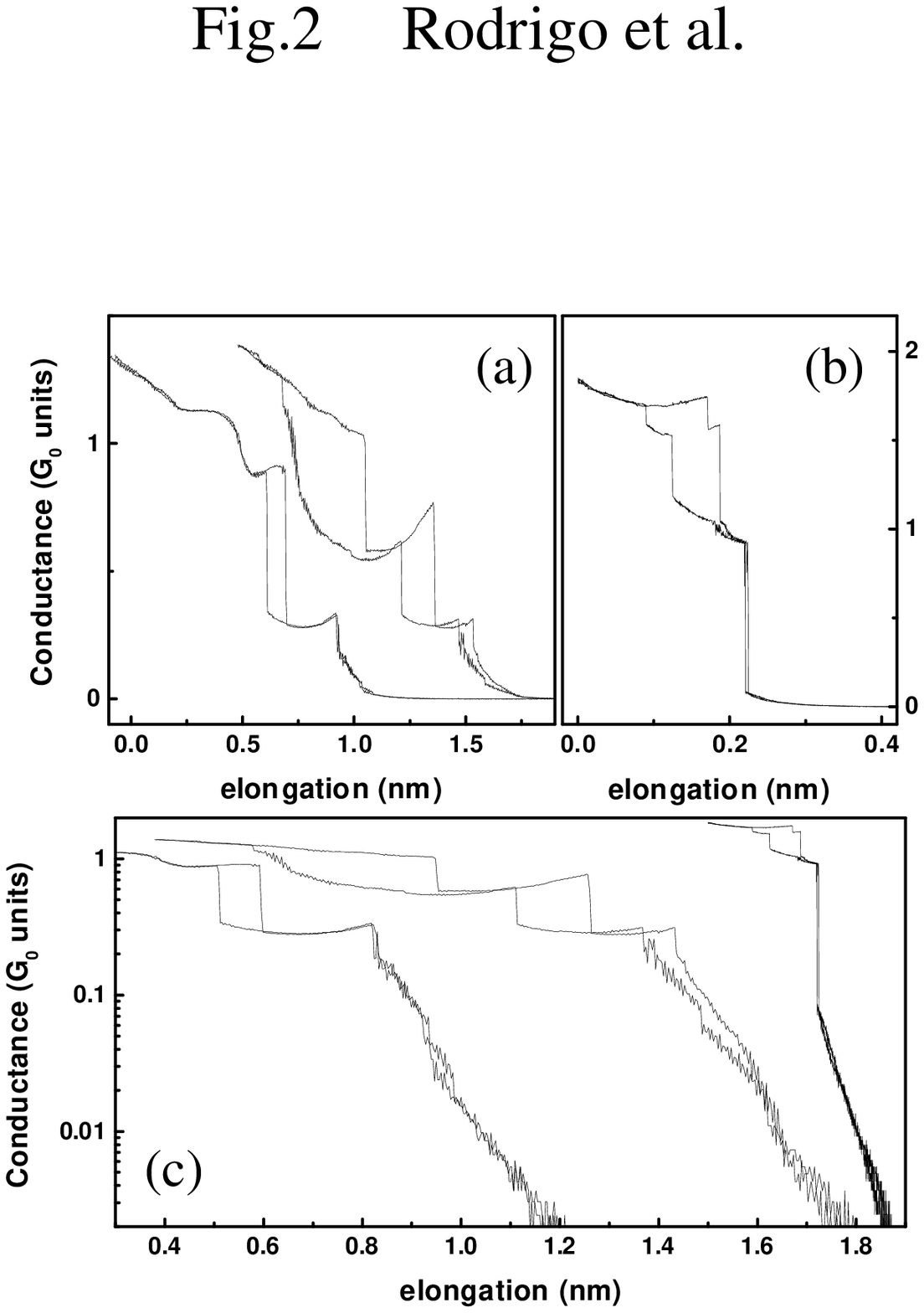,width=8cm,clip=}
\caption{Typical elongation-contraction cycles measured at 4 K: (a) and (b)
linear scale, (c)logarithmic scale. The cycles shown in (a) (the two cycles
on the left in (c)) present the sub-quantum conductance jumps and steps. The
cycle in (b) (the rightmost one in (c)) shows a contact breaking process.
Curves are shifted horizontally for clarity.}
\label{fig2}
\end{figure}

The experiments have been performed using a high stability scanning
tunneling microscope (STM) which can be operated over the range 2 K - 300
K inside a $^4$He cryostat. We have used high-purity (99.999\%)
polycrystalline bismuth, cut from a rod, for both STM tip and sample.
Prior to the measurements the surfaces were cut and scratched. At a given
location on the sample, the tip is crashed forcibly and repeatedly into
the substrate,  causing extended plastic deformations that lead to a
connective neck  between tip and sample. Different neck geometries can be
obtained by varying the indentation depth \cite{Agrait2}. In Fig.
\ref{fig1} (Fig. \ref{fig2} ) we show typical $G-Z$ curves corresponding
to a series of consecutive elongation cycles measured at 77K ( 4K ).
Continuous cycles of elongation-contraction of a neck lead to situations
in which the $G-Z$ curve presents clear steps over the entire conductance
range (typically from 0 to 10 G$_0$). Although the curves are not 
reproducible they show a very similar shape. Typically, the
length of the steps varies between 0.2 and 1 nm, depending on the initial
conditions of the elongation-contraction cycle (i.e., the neck)\cite{4K}.
These values are close to the lattice parameter of Bi (0.475nm). At 77 K
the conductance steps have a small positive slope without specific
structure. However, at 4K the first plateaus with $G\gtrsim G_0$ usually
show smooth curved shapes which resemble those observed for metallic Al
contacts \cite{Car-AlAu,uniaxial}. Moreover, the conductance
discontinuities $\Delta G$ are closer to $G_0$, 
suggesting that at 4K bismuth exhibits a metallic-like behaviour, with a
smaller $\lambda _F$. Therefore, only a few atoms are involved in the
contact. Distortions of the lattice parameter at the contact, together
with the peculiar band structure of bismuth, are responsible of
the curved plateaus obtained at 4 K.

As expected for a semimetallic contact, in the subquantum conductance
regime the $G-Z$ curves show clear steps at both temperatures (see Figs. 1
and 2). Except for the small jump discontinuities (caused by
rearrangements of the atoms and thus reflecting mechanical contact), $G$
decreases exponentially as a function of $Z$ like in a vacuum tunneling
process, but with an effective tunneling barrier that strongly depends on
temperature (see Figs. 1b and 2b). While at 77K the effective barriers are
typically of the order of 10meV, those obtained at 4K are close to 1eV.
When the contact finally breaks, the conductance is usually so small that,
due to limitations of our current experimental setup,  we are not able to
observe the transition from ``contact'' tunneling to actual vacuum
tunneling. All these facts indicate that, even at these very small values
of the conductance, tip and sample are still in mechanical contact. At 4K
there are some situations (see Fig. 2(b), and the curve on the right hand
side in Fig. 2(c)), in which the contact breaks before the subquantum
conductance regime. In those cases, the conductance does not display
subquantum steps and decreases exponentially as a function elongation,
where the effective tunneling barrier   $\phi \approx 4$eV is very close
to the expected bismuth work function.

\begin{figure}[]
\includegraphics[width=8cm]{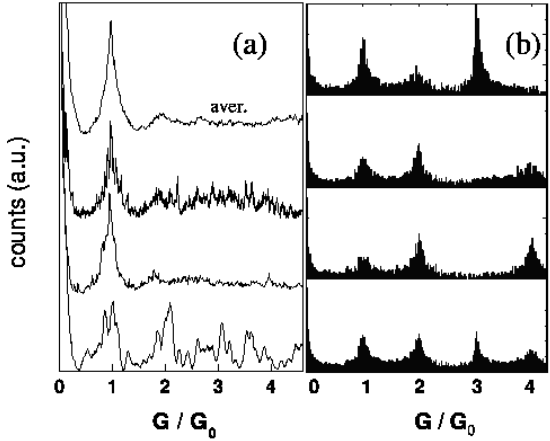}
\caption{Histograms of the conductance values obtained for different necks
at 77 K. (a) Experimental results from 200 $G-Z$ curves.  The top histogram
is an average of histograms corresponding to 10 different necks.
(b) Histograms obtained from the simulation of different initial necks. }
\label{fig3}
\end{figure}

A detailed theoretical description of the electrical properties of the
contact during elongation is a very difficult task. The effective mass of
the 
charge carriers in Bi varies over a wide range from $m^{*}\approx 0.002m_0$
for light electrons to $m^{*}\approx m_0$ for heavy electrons ($m_0$ is the
electron mass) and depends on the crystal orientation\cite{Bitheory,PRL2001}.
Moreover, the peculiar electronic properties of bismuth (semimetallic
character, band filling, Fermi surface, ...) may also vary with lattice
deformations due to the extreme uniaxial strains occurring in small contacts%
\cite{uniaxial} during the elongation-compression processes. The apparent
metallic behaviour at 4K can be qualitatively understood in terms of
quantum finite size effects \cite{rusosraros,Zhang}: as the contact radius $R
$ decreases the quantized transversal momentum $k_n\propto 1/R$ increases.
This leads to a shift in the subband energy position \cite{Zhang}, $%
E_n\approx \hbar ^2/(2m^{*})k_n^2$ (larger for light carriers) and the gap
between light and heavy electrons increases. At low temperatures, there is a
transition from light carriers at large cross sections to predominantly
heavy electron transport for small cross sections where the light electron 
states
become empty. This qualitative discussion is fully consistent with the
analysis of the tunneling regime in mechanical contact ($G<G_0$): The
effective barriers in mechanical contact are related to the electron
confinement in the constriction. In a free electron picture, the
longitudinal momentum of each transversal mode or channel $n$ is given by $%
k_{z_n}=\sqrt{K_F^2-k_n^2}$ (with $K_F=2\pi /\lambda _F$) such that to each channel 
$n$ corresponds a 1D effective potential barrier in the longitudinal
direction \cite{Torres94,Jacob55}. The modes with $k_n>K_F=(2\pi
/\lambda _F)$ are evanescent or tunneling modes. When $R\lesssim
\lambda _F$ there is no room even for a single channel and the conductance
is dominated by electron tunneling through the first channel with $%
k_{z_1}\approx ik_n\approx i\pi /R$. Assuming that the constriction length
increases linearly with the elongation $Z$ we will roughly have $G\propto
\exp (-\frac \pi RZ)$. Therefore, in the tunneling regime (before the contact
breaks), the slope of $\ln G$ as a function of $Z$ gives the
order of magnitude of the lateral size of the constriction. Our results then
suggest, that, at high temperatures, subquantum steps are related to structural
changes of constrictions involving hundreds of atoms ($R$ is a few
nanometers). Since the Fermi wavelength is larger or of the order of $R$,
it would also
be of the order of a few nanometers. In contrast, at low temperatures
subquantum conductance steps can be associated with changes in constrictions
consisting of a few atoms ($R$ is now a few \AA ). This would imply $\lambda _F$
to be one order of magnitude smaller than before. As mentioned above, this
qualitative discussion is fully consistent with the behaviour of the $G-Z$
curves in the range of a few conductance quanta.

\begin{figure}[]
\epsfig{file=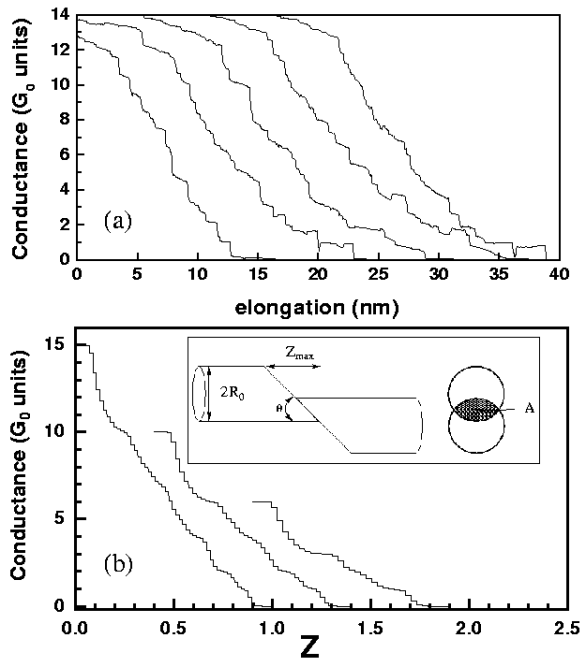,width=8cm,clip=}
\caption{(a) Experimental $G-Z$ curves at 77 K. (b) Simulated $G-Z$ curves
according to the model described in the text (different initial conditions). 
Inset: sketch of the deformation process. }
\label{fig4}
\end{figure}

The observation of subquantum steps in the $G-Z$
curves provides evidence of atomic rearrangements in the contact. Since
these curves are not exactly reproducible from cycle to cycle, 
conductance quantization effects are difficult to observe from a single elongation curve.   
After the formation of a given neck, a
histogram is obtained from 200 consecutive $G-Z$ curves. In fig. \ref{fig3}(a) we
present a set of  histograms obtained from different initial necks. As it
can be seen, the histograms
present peaks close to integer multiples of $G_0$ which, in contrast with
metallic contacts, can be unambiguously associated with conductance
quantization effects.  In general, we observe that the
histogram shape (peak height, number of peaks, ...) changes from neck to
neck \cite{4K}. This behaviour is clearly different than that observed in some
metals (e.g. gold nanocontacts \cite{Nanowires}). After averaging over 10 different
necks, the resulting histogram (uppermost in Fig. 3a) still shows clear peaks
at 1 and 2 conductance quanta which now resemble those of metallic
contacts. 
However, in semimetallic contacts a variation of a few quanta in the
conductance requires cross sections much larger than those in metallic
ones. Therefore, the mechanical evolution in these systems is expected
to be rather different than in metals. For
cross sections larger than a few nanometers, yielding occurs as a
consequence of collective shear events\cite{Stalder,Landman2} . A
qualitative picture of the mechanical evolution can be obtained from a
simple model system. We shall restrict ourselves to the tensile loading
regime during the elongation of the neck. Suppose that the initial
constriction is
represented by a cylinder of radius $R_0$ and consider the
deformation process sketched in the inset in Fig. \ref{fig4}(b). The contact is assumed
to glide along a symmetry plane that forms an angle $\theta $ with respect to
the cylinder axis. The maximum elongation $Z_{max}$ before the contact
breaks down, is given by $Z_{max}=2R_0/\tan (\theta )$. For a finite
elongation $\tilde Z=Z/Z_{max}$, the contact area $A$ shrinks according to (see inset Fig. 
\ref{fig4}(b)):
\begin{equation}
A(\tilde Z)=2R_0^2\left( \text{acos}(\tilde Z)-\tilde Z\sqrt{1-\tilde Z^2}%
\right) .
\end{equation}
Interestingly, the contact loses its axial symmetry and the narrowest
section of the contact resembles an elliptical cross section (Fig. \ref{fig4}%
(b)) with an axis aspect ratio $\eta (\tilde Z)\approx \pi R_0^2(1-\tilde Z%
)^2/A$. An estimation of the transmission coefficients $T_{nm}$ as a
function
contact elongation can be obtained from a simple model based on the
saddle-point-contact model \cite{Buttiker,Tito}. The geometry of our contact
resembles the wide-narrow-wide geometry with `elliptical' cross section
discussed in Ref. \cite{Tito}. Therefore, $T_{nm}$ can be approximated by: 
\begin{equation}
T_{nm}=\left[ 1+\exp \left( \frac{-2\pi (\sqrt{\pi A}-\epsilon _{nm})}{\tan
\alpha }\right) \right] ^{-1}
\end{equation}
where $\epsilon _{nm}=(n+1/2)/\sqrt{\eta }+\sqrt{\eta }(m+1/2)$ (all the
lengths are in units of $\lambda _F$), and $\tan \alpha =\sqrt{(A(0)-A(\tilde Z%
))/\pi }/Z_{max}$ defines the effective ``opening angle'' \cite{Torres94} of the
constriction. The conductance curves are obtained from the Landauer formula, 
$G=\sum_{nm}T_{nm}$.

Typical $G-Z$ curves obtained by this simple model for $\theta = 50^\circ$
are shown in Fig. \ref{fig4} together with some experimental results at
77K. In order to mimic the stepped structure, the cross section is assumed
to be constant within a lattice parameter of Bi. Experimentally the conductance
changes from $\approx 10 G_0$ to $G_0$ after an elongation of the contact
of the order of 15 nm. Within our 
model this fixes the Fermi wavelength to be in the nanometer scale, in
agreement with the analysis of the subquantum regime. For small $Z$, i.e.
for
small opening angles, there is almost no scattering (the $T_{nm}$'s are
0 or 1). Therefore, the conductance is simply given by the number of propagating
channels in the initial contact $N_0$, i.e., those $nm$ channels having $%
\epsilon_{nm} < \sqrt{\pi A}$. As the elongation proceeds  an
interesting interplay between the narrowing of the constriction and the
change of degeneracy of transversal modes can be observed. 
For $\tilde{Z} = 0$ the contact has cylindrical symmetry and the
transversal levels exhibit the well known degeneracy of a parabolic
confining potential. Depending on the initial contact cross section, the conductance 
is then given by $1,3,6,10, \dots G_0$. As $\tilde{Z}$ increases, the
cross section decreases and, at the same time, the degree of anisotropy $%
\eta(\tilde{Z})$ also varies, removing the mentioned degeneracy. The
conductance behaviour along the elongation path depends critically on the
initial conditions of the contact. The theoretical histograms (Fig. \ref
{fig3}(b)) strongly depend on the initial cross section and exhibit an
overall shape that is very similar to the experimental one.

In summary, the above experiments provide direct experimental evidence for
the presence of conductance quantization in three-dimensional nanocontacts, 
as well as an experimental exploration of the conductance regime
prior to the opening of the first quantum channel. Our results for
moderate-sized
contacts are well described by a simple model of contact evolution during an
elongation process that is based on a gliding mechanism. In agreement with
magnetotransport measurements, the differences observed between 77K and 4K
curves for $G\lesssim G_0$ are consistent with a transition from light to
heavy electrons at the Fermi level when the contact cross section becomes
sufficiently small.  
  Finally, we consider that our work opens a new way to perform
 experiments on quantum coherence phenomena  in atomic scale
  contacts that up to now were doable with semiconducting
  heterostructures only.

This research was supported by the DGICyT (Spain) through grants
BFM2000-1470-C02-02, PB98-0464 and PB97-0068.

\narrowtext
\end{multicols}


\begin{references}
\bibitem[\dagger]{adress}  Present address: Institut f\"ur Theorie der
Kondensierten Materie, Universit\"at Karlsruhe, P.O. Box 6980, 76128
Karlsruhe, Germany

\bibitem{Muller}  C.J. Muller, J.M. van Ruitenbeek and L.J. de Jongh, Phys.
Rev. Lett. {\bf 69}, 140 (1992); J.M. Krans {\em et al.}, Nature {\bf 375},
767 (1995).

\bibitem{Agrait}  N. Agra\"\i t, J.G. Rodrigo and S. Vieira, Phys. Rev. B 
{\bf 47}, 12345 (1993); J.I. Pascual {\em et al.}, Phys. Rev. Lett. {\bf 71}%
, 1852 (1993); L. Olesen {\em et al.}, Phys. Rev. Lett. {\bf 72}, 2251
(1994).

\bibitem{Costa1}  J.L. Costa-Kr\"amer, N. Garc\'\i a and P. Garc\'\i
a-Mochales and P.A. Serena, Surf. Sci. Lett. {\bf 342}, L1144 (1995).

\bibitem{Nanowires}  P.A. Serena and N. Garc\'\i a (eds.), ``{\em Nanowires}%
'', NATO-ASI Series E: Appl. Sci. {\bf 340} (Kluwer, Dordrecht, 1997); J.M.
van Ruitembeek in ``{\em Mesoscopic Electron Transport}'', ed. by L.L. Shon,
L.P. Kouwenhoven and G. Sch\"on, NATO-ASI Series E: Appl. Sci. {\bf 345}
(Kluwer, Dordrecht, 1997) p549-579.

\bibitem{channels}  E. Scheer {\em et al.}, \prl  {\bf 78,} 3535 (1997);
Nature {\bf 394}, 154 (1998).

\bibitem{Landman}  U. Landman, W.D. Luedtke, N.A. Burnham and R.J. Colton,
Science {\bf 248}, 454 (1990); T.N. Todorov and A.P. Sutton, Phys. Rev.
Lett. {\bf 70}, 2138 (1993); J.A. Torres and J.J. S\'aenz, \prl {\bf 77},
2245 (1996).

\bibitem{Agrait2}  N. Agra\"\i t {\em et al.}, Thin Solid Films {\bf 253},
199, (1994); Phys. Rev. Lett. {\bf 74}, 3995 (1995); G. Rubio, N. Agra\"\i t
and S. Vieira, \prl {\bf 76}, 2302 (1996); C. Untiedt, G. Rubio, S. Vieira
and N. Agra\"\i t, \prb {\bf 56}, 2154 (1997).

\bibitem{Krans-Sb}  J. M. Krans, and J. M. van Ruitenbeek, Phys. Rev. B {\bf %
50}, 17659 (1994).

\bibitem{Torres94}  J.A.Torres, J.I. Pascual and J.J. S\'aenz, Phys. Rev. B 
{\bf 49}, 16581 (1994).

\bibitem{Olin}  J.L. Costa-Kr\"amer, N. Garc\'\i a and H. Olin, \prl {\bf 78}%
, 4990 (1997).

\bibitem{Pethica}  J.K. Gimzewski and R. M\"oller, \prb {\bf 36}, 1284
(1987).

\bibitem{Bitheory}  S. Golin, Phys. Rev. {\bf 166}, 643 (1968); R.D. Brown,
R.L. Hartman and S.H. Koenig, Phys. Rev. {\bf 172}, 598 (1968).                          

\bibitem{PRL2001} C.R. Ast and H. Hoechst, Phys. Rev. Lett. 87, 177602
(2001)

\bibitem{Zhang}  Z. Zhang, X. Sun, M.S. Dresselhaus and J.Y. Ying, \apl {\bf %
73}, 1589 (1998).

\bibitem{4K}  At 4K, even for apparently similar initial conditions, the $G-Z
$ curves may differ considerably between each other. Sometimes the step
lengths could be extremelly large compared with the lattice parameter thus
leading to spurious peaks in the conductance histograms. Clear peaks close
to integer multiples of $G_0$ could only be obtained from selected curves.
An averaged histogram over many different initial necks shows no
 defined peaks at any conductance value.

\bibitem{Car-AlAu}  J. M. Krans {\em et al.}, \prb {\bf 48}, 14721 (1993);
J. C. Cuevas {\em et al.}, \prl  {\bf 81}, 2990 (1998).

\bibitem{uniaxial}  D. S\'anchez-Portal {\em et al.}, \prl {\bf 79}, 4198
(1997).

\bibitem{rusosraros}  V.N. Lutskii, JETP Letters {\bf 2}, 391 (1965).

\bibitem{Jacob55}  M. Brandbyge, K.W. Jacobsen and J.K. N{\o }rskov, \prb 
{\bf 55}, 2637 (1997).

\bibitem{Stalder}  A. Stalder and U. D\"urig, J. Vac. Sci. Technol. B {\bf 14%
}, 1259 (1996); Appl. Phys. Lett. {\bf 68}, 637 (1996).

\bibitem{Landman2}  U. Landman, W.D. Luedtke, B.E. Salisbury and R.L.
Whetten, \prl {\bf 77}, 1362 (1996).

\bibitem{Buttiker}  M. B\"uttiker, \prb {\bf 41}, 7906 (1990); A.G.
Scherbakov, E.N. Bogachek and U. Landman, \prb {\bf 53}, 4054 (1996).

\bibitem{Tito}  T. L\'opez-Ciudad {\em et al.}, Surf. Sci. Lett {\bf 440},
L887 (1999).

%%%%%%%%%%%%%%%%%%%%%%%%%%%%%%%%%%%%%%%%%%%%%%%%%%%%%%%%%%%%%%%%%%%%%%%%%%%
\end{references}
\end{document}